\documentclass[10pt,letterpaper]{article}
\usepackage{opex3}
\usepackage{amsmath,amssymb}

\begin{document}

\title{Phase-locking of two self-seeded tapered amplifier lasers}

\author{G.~Tackmann*, M.~Gilowski*, Ch.~Schubert, P.~Berg, T.~Wendrich, W.~Ertmer and E.~M.~Rasel}

\address{Institut f{\"u}r Quantenoptik\\ Gottfried Wilhelm Leibniz Universit{\"a}t Hannover\\ Welfengarten 1\\ 30167
Hanover, Germany\\ * Co-first authors}

\email{tackmann@iqo.uni-hannover.de} 



\begin{abstract}
We report on the phase-locking of two diode lasers based on self-seeded tapered amplifiers. In these lasers, a reduction of linewidth is achieved using narrow-band high-transmission interference filters for frequency selection. The lasers combine a compact design with a Lorentzian linewidth below 200~kHz at an output power of 300~mW for a wavelength of 780~nm. We characterize the phase noise of the phase-locked laser system and study its potential for coherent beam-splitting in atom interferometers.
\end{abstract}




\section{Introduction}

Phase-locked diode laser systems have become an
important instrument for coherent beam-splitting in Raman type atom interferometers~\cite{Santaelli94},~\cite{Berman97}. Providing high precision, these atom interferometers are, for example, used to measure the gravity
constant~\cite{Lamporesi08}, Earth's gravity~\cite{Gouet08_1},
or to test the equivalence principle~\cite{Dimopoulos07}. The high
precision sensing of rotation and acceleration with stationary or
transportable atom interferometers is a growing field of actual
research~\cite{Finaqs}. These applications, being the motivation for the work presented in this article, impose stringent requirements on the properties of the laser systems. They have to provide, on the one hand, a low phase noise. Recent efforts in this direction have been made, for instance, by the implementation of an intra-cavity electro-optical modulator increasing the servo bandwidth of
the phase-locked loop~\cite{Gouet08_2}. On the other hand, the output power of the lasers should be high enough in order to enable short interaction times between the light fields and the atoms during the beam splitting process. Thus, a larger number of atoms can be addressed in a velocity selective Raman transition and higher pulse efficiencies can be reached. In this way, the signal-to-noise ratio and the contrast can be augmented, leading to a better sensitivity of the atom interferometer. Moreover, lasers with high output powers enable multi-photon transitions, which, in principle, allow to further improve the sensitivity of atom interferometers~\cite{HMueller08}. These optical powers have been realized by, for example, operating tapered amplifiers in a constant seed power mode but with a pulsed current
supply~\cite{Takase07}. Finally, atom interferometers proposed for space
missions~\cite{Rasel} or drop-tower experiments~\cite{Quantus}, as
well as transportable sensors~\cite{Finaqs}, require compact laser
set\-ups. These applications challenge todays phase-locked laser systems such as master oscillator power amplifier (MOPA) systems using diode lasers~\cite{Wilson98} or Titanium Sapphire lasers~\cite{HMueller08}.

In the this article, we present and characterize a
phase-locked laser system based on self-seeded tapered amplifiers
using narrow-band high-transmission interference filters
for frequency selection. Our technique is very much in line with
the recent progress in the development of novel phase-locked diode
laser systems, since the realized system comprises high output
power and compactness. We studied the phase noise of the laser
system and tested its potential for the beam splitting process
in atom interferometry.

\section{Laser and phase-locked loop setup}
Before describing the setup of the optical phase-locked loop, we
briefly introduce the design of the used tapered lasers as well as
their characteristics, since a detailed description is presented
in~\cite{Gilowski07}.

\begin{figure}[htbp]
\centering\includegraphics[width=10cm]{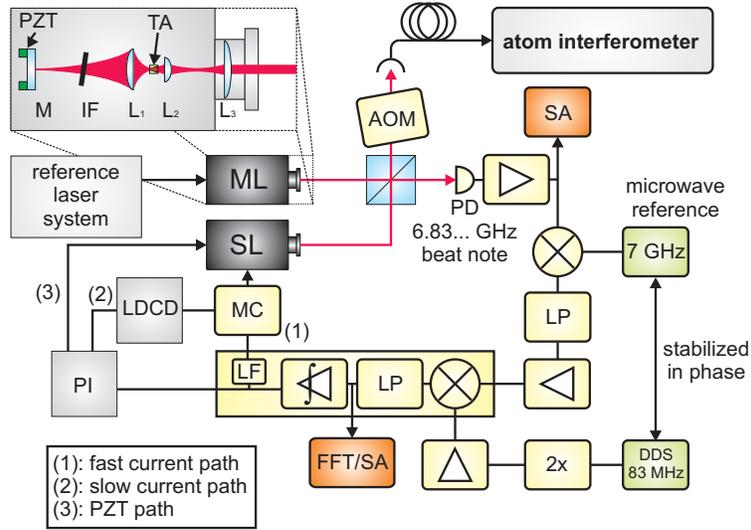}
\caption{Phase-locked loop setup. The beat note of the two lasers is
detected by a photodiode and a control signal is generated by the
subsequent elements, which is fed back to the slave laser via
three control loops. LP and LF stand for low-pass-filter and loop-filter, respectively. The other acronyms can be found in the text.} \label{fig:PLLScheme}
\end{figure}

As shown in Fig.~\ref{fig:PLLScheme}, we utilize two lasers, a
master laser (ML) and a slave laser (SL), each comprising a tapered
amplifier laser diode (TA)~\cite{EagleYard}, which emits light in two opposite
directions. The rear facet of the tapered amplifier is AR coated,
whereas the front facet has a low reflectivity. The latter one forms,
together with a HR coated mirror (M), the cavity with a length of 41~mm. The mirror is
mounted on a piezoelectric transducer (PZT) for controlling the cavity length.

With the aid of the intra-cavity lens $\textnormal{L}_1$, light coming from the
rear facet of the TA is focused onto the mirror, backreflected into
the diode and thus realizing the self-seeding of the TA.
We expected, that the optical feedback will be influenced by the numerical aperture of this lens. Therefore, we tested different values and by using aspheric lenses of $f=4.5~\textnormal{mm}$ and $NA=0.55$ in the case of the ML
and of $f=3.1~\textnormal{mm}$ and $NA=0.68$ for the SL. The resulting output power is presented below. The frequency
selection is performed by a narrow-bandwidth, high-transmission
interference filter (IF)~\cite{Baillard06,Gilowski07} located
between the mirror and lens $\textnormal{L}_1$. This filter is adjustable in its angle with
respect to the propagation axis of the beam. The output beam is
collimated by a compact lens design consisting of an aspheric lens
$\textnormal{L}_2$ with $f=2~\textnormal{mm}$, $NA=0.50$ and a cylindrical lens $\textnormal{L}_3$ with a
focal length of $f=15~\textnormal{mm}$~\cite{Takase07}. With this compact
collimation setup we obtain injection efficiencies into a
single-mode optical fiber of up to 60\%.

We obtain an output power of $P^{ML}_{out}=763~\textnormal{mW}$ and $P^{SL}_{out}=1084~\textnormal{mW}$ before passing an optical isolator at a supply current of $1.9$~A and $2.0$~A, respectively, while the threshold currents are $I^{ML}_{th}=970~\textnormal{mA}$ and $I^{SL}_{th}=930~\textnormal{mA}$. This confirms the dependence of the optical output power on the numerical aperture of the intra-cavity lens. A higher numerical aperture provides more optical feedback, and this way results in a higher optical output power and lower threshold current.

For powers above 500~mW, we observe spatial multi-mode operation. We attribute this effect to light field propagations in the TA chip extending out of the waveguide structure as is described in~\cite{FBI2}. Here, the instabilities of the spatial mode have been suppressed by adding reverse bias absorber sections adjacent to the ridge waveguide of the TA chip. Replacing the TA chip used in our lasers by one featuring reverse bias absorber sections can suppress the multi-mode operation for high output powers.

To avoid multi-mode operation, the linewidths of the two lasers are measured at powers of less than 300~mW, where a single-mode operation is assured. To determine these linewidths, we measure the beat note between each of the lasers and a narrow linewidth external cavity diode laser (ECDL) which is stabilized in frequency onto an atomic transition during the beat note measurements. From a set of 20 shots, measured with a sampling time of 28~ms and a resolution bandwidth of 30~kHz, we find a broadened linewidth of the ML of 494~kHz and 457~kHz for the SL, as well as a Lorentzian linewidth of 195~kHz and 190~kHz.

For the realization of an optical phase-locked loop, the lasers are operated at an output power of $P^{ML}_{out}=223~\textnormal{mW}$ and $P^{SL}_{out}=289~\textnormal{mW}$, respectively. We implemented the scheme shown in Fig.~\ref{fig:PLLScheme}. In this, the ML is stabilized in its
frequency to the above mentioned ECDL which is the reference laser
in our experiment. The light of the ML and SL is superimposed by a
beam splitter (BS). The first output port leads to an acousto-optical modulator (AOM) and, subsequently, to a single-mode fiber guiding the light to the atom interferometer. On the second port, the beat note signal at $\omega_0=2\pi\times6.83~\textnormal{GHz}$, the frequency of the ground state Hyperfine
transition in $^{87}\textnormal{Rb}$, is detected by a photodiode (PD). This signal is downconverted in two steps. First, a signal of 167.5~MHz is produced by mixing the beat note signal with the 7~GHz output of an ultra-stable microwave reference analogous to~\cite{Cheinet08}. The obtained signal is then downconverted to a DC signal by mixing it with a sinusoidal of the same frequency, which is provided by a frequency doubled 83.75~MHz signal of a direct digital synthesizer (DDS). Together with a low pass filter, the mixer used for this second downconversion forms an analog phase detector generating the phase error signal.

The control signal, generated out of the error signal by a first amplification stage, is fed back to the SL via three paths. A feed back loop for high frequencies (1) is realized by directly modulating the diode current using a special Bias-T (MC), which is able to handle DC currents up to 3~A as required for operating the tapered amplifier. For the medium (2) and low frequency regime (3), the control signal is firstly integrated by a second PI controller (PI) and then controlling the current set point of the laser diode current driver (LDCD) and the PZT voltage. In this way, we realize a stabilization of the phase difference of the two lasers onto a very stable, low noise microwave reference.

\section{Characterization of phase noise and its impact on the atom interferometer}
\label{sec:characterization}

We analyze the phase noise of the laser system via two methods~(see Fig.~\ref{fig:PLLScheme}). First, we record the beat note at 6.83~GHz with a spectrum analyzer (SA) after the first amplification stage. Second, the
in-loop power spectral density $S_{\varphi}$ of the
phase noise is directly measured by recording the error signal
of the analog phase detector. The
frequency range up to 100~kHz is measured with a spectrum analyzer
providing a fast Fourier transformation routine (FFT), while
higher frequency contributions to the phase density are recorded
with a usual spectrum analyzer (SA) device.

\begin{figure}[htbp]
\centering\includegraphics[width=8.5cm]{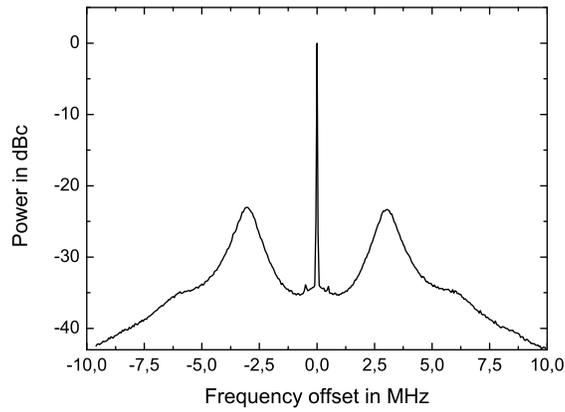}
\caption{Beat note of the two phase-locked lasers measured at a
frequency of 6.83~GHz. The graph shows an average over
100~measurements with a sweep time of  317.3~ms and a resolution
bandwidth of 10~kHz.} \label{fig:beatnote}
\end{figure}

\begin{figure}[bp]
\centering\includegraphics[width=8.5cm]{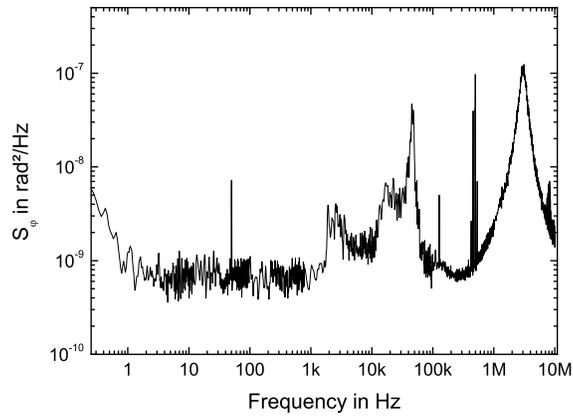}
\caption{In-loop power spectral density of the phase noise in the
phase-locked loop. The data points were taken in several steps according
to the frequency range with an average of 20 measurements.}
\label{fig:FFT_PSD}
\end{figure}

Closing all feedback loops, we obtain the beat note shown in
Fig.~\ref{fig:beatnote}. The PLL bandwidth is 3~MHz and the noise
background is suppressed by about 35~dB relative to the carrier. The phase noise spectral density, measured after the analog phase
detector, is displayed in Fig.~\ref{fig:FFT_PSD}. In the frequency
range from 2~Hz to 2~kHz, the power spectral density has a plateau
at about $6.5\times10^{-9}~\textnormal{rad}^2/\textnormal{Hz}$. Above, the power spectral density increases at 2~kHz, 40~kHz and 3~MHz. These maxima originate from the limited servo bandwidth of the PZT, the slow and the fast current feed back paths, respectively. By integrating $S_{\varphi}$ from 2~Hz up to 10~MHz and taking the square root of the result we find a phase noise of $\sigma_{\varphi}=~907~\textnormal{mrad}$.

\begin{figure}[bp]
\centering\includegraphics[width=10cm]{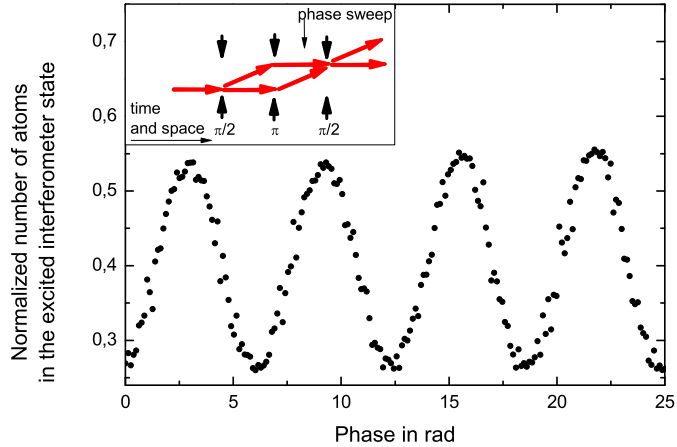}\caption{Interference fringes of the described atom interferometer
configuration. The inset illustrates the three pulse sequence and
the resulting atomic trajectories. To observe the fringes, the
phase difference between the two phase-locked lasers is scanned
before the last interferometry pulse.} \label{fig:MZfringes}
\end{figure}

With respect to our application, we study the phase noise contribution of the laser system employed for coherent beam splitting in our Raman type atom interferometer. The latter is described in detail in~\cite{TMueller09}. In this, laser-cooled Rubidium~87 atoms are coherently split, reflected and recombined in a $\pi/2$-$\pi$-$\pi/2$-configuration~\cite{Borde89} by the use of a 2-photon Raman transition between the two Hyperfine states of the electronic ground state (see the inset of Fig.~\ref{fig:MZfringes}).

The phase noise of the laser system enters into the signal of the atom interferometer via the beam splitting process. In order to determine the phase noise contribution of the laser system, the power spectral density of the phase-locked loop's phase noise has to be weighted with the weighting function derived from the sensitivity function of the atom interferometer. Similar to the description of fountain clocks, the sensitivity function considers the pulsed mode operation of the atom interferometer. Since this formalism is extensively described in~\cite{Cheinet08}, we briefly present it here.

To obtain the impact of the laser phase noise onto the total phase noise of the interferometer, the spectral phase density $S_{\varphi}$ has to be weighted. Thus, the total phase noise per interferometer shot can be derived from
\begin{equation}\label{eq1}
\sigma_{\Phi}^2=\int_{0}^{\infty}|H(f)|^{2}S_{\varphi}(f)\textnormal{d}f,
\end{equation}
where the weighting function
\begin{eqnarray}\label{eq2}
H(f)&=&4i\pi f\frac{\sin\left[\pi f\left( T+\tau\right)\right]\Omega_R\left\{ 2\pi f\cos\left[\pi f\left( T+\tau\right)\right]+\sin\left[\pi fT\right]\Omega_R\right\}}{4\pi^3f^3-\pi f\Omega_R^2}
\end{eqnarray}
is given by the Fourier transform of the function $h(t)$ and works as a low pass filter for the spectral phase noise density. In Eq.~(\ref{eq2}), $\Omega_R$ is the Rabi frequency, $\tau$ is the $\pi$-pulse length and $T$ is the time between to pulses. The function $h(t)$ describes the sensitivity of the interferometer for small phase jumps imprinted during the three pulse sequence~\cite{Cheinet08}. Using Eq.~(\ref{eq1}), we calculate the total phase noise contribution of the laser system to the inter\-fero\-meter per shot to be $\sigma_{\Phi}=140~\textnormal{mrad}$.

In order to determine the total phase noise contribution
experimentally, we measure atomic interference fringes while changing the phase difference of the two
lasers stepwise for each cycle. This can be performed by sweeping the phase
offset of the DDS before applying the third beam splitting pulse
(see inset of Fig.~\ref{fig:MZfringes}). The resulting
interferometer fringes are shown in Fig.~\ref{fig:MZfringes}.
In these measurements, the time between two beam
splitting pulses is $T=0.5~\textnormal{ms}$, and the length of the
$\pi$-pulse is $\tau=11~\mu\textnormal{s}$. From the fringe pattern we
infer a contrast of 35\% as well as a signal-to-noise ratio
of 7 by analyzing the amplitude noise on the pattern's edges
which leads to a total phase noise of $144\pm 8~\textnormal{mrad}$. A part of this noise is due to vibrational and detection noise in the atom interferometer. We can derive an upper limit for these noise contributions by analyzing a measurement, which has been performed with the same experimental parameters, but with a low noise laser system. The total phase noise in this measurement is 24~mrad. Combining this value with the noise derived from the sensitivity function formalism, we obtain 142~mrad as the total phase noise value, which agrees well with the measurement.

Indeed, the phase noise of the presented laser system is bigger
then in other phase-locked diode laser systems such as
in~\cite{Gouet08_2}. We attribute this to the spectral properties
of the tapered amplifier diode. Since its full width half maximum
linewidth is 26~nm and thus about 4 orders of magnitude broader
than the linewidth of a common laser diode, the demand for the
optical feedback and the electronic feedback loop is much higher
in the case of a TL than in a common ECDL system. The spectral
noise floor of a free running TL which can be inferred from a beat
measurement is at about -40~dBc and has to be
compared to a spectral noise floor of {-60} to {-70}~dBc in the case
of a common ECDL system~\cite{Gilowski07}.

There are different solutions to improve the performance of the laser system regarding the phase noise. For this, the white phase noise can be reduced by increasing the frequency selectivity of the resonator. This can be achieved, either by expanding the resonator length, or via an increase of the reflectivity of the tapered amplifier's front facet in combination with the implementation of reverse bias absorber sections. Furthermore, the noise floor can be reduced by coupling light into a cavity and providing additional optical feedback to the tapered laser~\cite{Wicht07}. Apart from this, the use of an improved current source can further diminish the white noise of the tapered lasers. The sources, which are used for the measurements presented here, provide a DC current of up to 2~A and a rms noise of $15~\mu\textnormal{A}$.

\section{Conclusion}

In this work, we reported on the first phase-locking of two self-seeded tapered amplifier lasers using an interference filter for frequency selection. Providing a single-mode optical output power of up to 300~mW each, the two tapered lasers have a free running Lorentzian linewidth of less than 200~kHz. The single-mode optical output power of the laser system can, in principle, be increased up to 1~W per laser by adding reverse bas absorber sections to the TA chip in order to suppress spatial multi-mode operation. The phase-locked loop, realized for more than 200~mW optical power per laser, provides a servo bandwidth of 3~MHz and a total phase noise of 907~mrad in the frequency range from 2~Hz to 10~MHz. Employing this phase-locked laser system for Raman beam splitting in our atom interferometer, we find a total interferometer phase noise of $144\pm 8~\textnormal{mrad}$. By weighting the in-loop power spectral density of the phase noise with the sensitivity function of the atom interferometer, we calculate a phase noise contribution of the laser system to the atom interferometer phase noise of 140~mrad. Deducing an upper limit for other noise sources in the atom interferometer, we find a good agreement of the laser system's phase noise contribution with the calculated value.

The laser system represents an attractive source of Raman beam-splitter light fields for applications such as in~\cite{leveque09}. Here, the influence of the laser phase noise is strongly suppressed. Furthermore, a reduction of the phase noise of the laser system can be achieved by exploiting the improvements described in section~\ref{sec:characterization} for making it more suitable for state-of-the-art Raman type atom interferometers.

Considering its compactness, our laser system is particularly attractive for atom interferometer experiments, in which compactness is an important issue. These are, for example, transportable sensors, airborne devices, or experiments in microgravity environments. In this way, our work opens up new vistas for the miniaturization of laser sources for coherent beam splitting for atom interferometers. Further progress in this can be achieved by implementing the presented external cavity design comprising an interference filter for frequency selection on a micro-optical bench~\cite{FBI}.

\section{Acknowledgements}
This work was supported in part by the Deutsche
Forschungsgemeinschaft (SFB 407) and in part by the European Union
(Contr. No. 012986-2 (NEST), FINAQS, Euroquasar, IQS). We also thank the center for
quantum engineering and space time research QUEST. M.G. and G.T. would like
to thank the Max-Planck-Gesellschaft for financial support. Additionally,
we would like to thank Christian Fiebig from the Ferdinand-Braun-Institut
f\"ur H\"ochstfrequenztechnik for fruitfull discussions.

\end{document}